\documentclass[aps,prc,preprint,showpacs,preprintnumbers,amsmath,amssymb] {revtex4-1}
\usepackage{graphicx}
\usepackage{amsmath}
\usepackage{mathrsfs}
\usepackage{textcomp}
\usepackage[T1]{fontenc}
\usepackage{lmodern}
\renewcommand{\vec}[1]{\boldsymbol{#1}}

\newcommand{\La}{\ensuremath{\mathcal L}}

\def\be{\begin{equation}}
\def\ee{\end{equation}}
\def\bg{\begin{eqnarray}}
\def\en{\end{eqnarray}}
\def\nn{\nonumber}

\usepackage{color}

\begin{document}
\preprint{\bf LFTC-18-17/38}

\title{Description of the recently observed hypernucleus 
$^{15}{\!\!\!_{\Xi^-}}$C within a quark-meson coupling model} 
\author{R. Shyam$^{1}$, K. Tsushima$^2$}
\affiliation{$^1$ Theory Division, Saha Institute of Nuclear Physics, 
1/AF Bidhan Nagar, Kolkata 700064, India \\
$^2$ Laborat\'{o}rio de F\'{i}sica Te\'{o}rica e Computacional, 
Universidade Cruzeiro do Sul, Rua Galv\~{a}o Bueno, 868, Liberdade 
01506-000, S\~{a}o Paulo, SP, Brazil } 

\date{\today}
\begin{abstract}
We investigate within a quark-meson coupling (QMC) model, the  
structure of the  bound $\Xi^-$ hypernucleus 
$^{15}{\!\!\!_{\Xi^-}}$C ($^{14}$N + $\Xi^-$), which has been 
observed in a recent analysis of the KEK-E373 experiment. In the 
QMC model, light quarks in nonoverlapping nucleon and  
$\Xi^-$ bags interact self-consistently with isoscalar-scalar 
($\sigma$), isoscalar-vector ($\omega$), and isovector-vector 
($\rho$) mesons in the mean field approximation. The parameters of 
the model (quark-meson coupling constants and masses) are mostly 
fixed from the nuclear matter saturation properties. The QMC model 
closely reproduces the separation energies of the two $\Xi^-$ 
hyperon states in $^{15}{\!\!\!_{\Xi^-}}$C reported in the 
KEK-E373 analysis, and identifies their quantum numbers. We also 
make predictions for the cross sections for the production of the 
$^{15}{\!\!\!_{\Xi^-}}$C hypernuclear specta in the ($K^-, K^+$) 
reaction on a $^{15}$O target within a covariant effective 
Lagrangian model using the $\Xi^-$ bound state spinors obtained 
within the same QMC model.

\end{abstract}
 
\maketitle

\newpage

The knowledge about the hyperon-nucleon ($YN$) and hyperon-hyperon
($YY$) interactions is still not of the same precision as it is 
for the nucleon-nucleon ($NN$) interaction. This is mainly due to 
the fact that sufficient and precise scattering data on the $YN$ 
and $YY$ systems are not available because such experiments are 
difficult to perform due to the short life times of hyperons. This 
is of vital importance to have precise information about these 
interactions as they are key to the investigations of several 
interesting aspects in nuclear and astrophysics. This input is 
needed in probing the role of the strangeness in the equation of 
state at high density, as investigated in the cores of neutron 
stars~\cite{bie10,lat04} and in high energy heavy ion collisions 
at relativistic heavy ion collider (RHIC) located in Brookhaven 
National laboratory~\cite{ada07}, CERN~\cite{aam11} in Geneva and 
FAIR facility in GSI~\cite{CBM11}, Darmstadt.

During the last decade lattice simulations, which perform the {\it 
ab initio} calculations of quantum chromodynamics (QCD) in the 
nonperturbative regime, have been developed to calculate 
baryon-baryon interactions in both nonstrange~\cite{ish07,aok10,
ish12} and strange sectors~\cite{nem09,ino10,ino11,nem18,sas18} by 
the HAL QCD collaboration. However, more studies are needed with 
improved statistics at the physical pion mass point to stabilize 
the results of these investigations. 

The studies of hypernuclei provide another means to get useful 
information about the $YN$ interaction. For the $\Lambda$ 
(strangeness $S$ = -1) hypernuclei extensive amount of experimental 
data are available for some of the structure observables (like 
binding energies and spins)~\cite{has06,fel15}. This has provided 
important constraints (see, e.g.~\cite{gal16} for a recent review) 
on the bare $\Lambda N$ interactions~\cite{nag79,mae89,reu94,fuj98,
rij99,hai05,hai13} obtained by fitting the sparsely available 
$\Lambda N$ scattering data. 

As far as $\Xi$ hypernuclei are concerned, there are some hints of 
their existence from emulsion events~\cite{wil59,bec68,aok93,aok95,
yam01}. The $(K^-, K^+)$ reaction that leads to the transfer of two 
units of both charge and strangeness to the target nucleus, provides 
one of the most promising ways for studying the $\Xi$ hypernuclei. 
However, in a few experiments performed for this reaction on a 
$^{12}$C target~\cite{fuk98,kha00}, no bound state of the 
hypernucleus $^{12}{\!\!\!_{\Xi^-}}$Be was unambiguously observed 
because of the limited statistics and detector resolution. 
Nevertheless, from the studies on the shapes of the continuum 
spectra observed in these experiments an attractive $\Xi^-$ - 
nucleus potential well depth of about 14 MeV was suggested. On the 
other hand, in Ref.~\cite{gal83}, a  value in the range of  21$-$24 
MeV was obtained for the depth of this potential well from the 
analysis of scarce emulsion data. At the same time,  in a recent 
study~\cite{shy12} performed for the $^{12}{\!\!\!_{\Xi^-}}$Be 
hypernucleus within a version of the QMC model~\cite{gui08} that 
is constructed for applications to the strange hyperon sector, the 
sum of the scalar and vector mean field potentials at the zero 
radius was found to be about -21.3 MeV. This quantity, which is 
comparable with the depth of the conventional Woods-Saxon potential
\cite{shy12,shy13,tsu13}, is similar to the $\Xi^-$-nucleus 
potential depth obtained in the analysis of Ref.~\cite{gal83}. 
Thus, the $\Xi^-$-nucleus potential depths obtained in analyses of 
emulsion and spectrometer data and in alternative theoretical 
models differ from each other by as much as 10 MeV.

Convincing evidence for the $\Xi$ single-particle bound states is
of crucial importance for the extraction of the reliable information
on the $\Xi$ single-particle potential and the effective $\Xi N$
interaction. Recently, in Ref.~\cite{nak15}, the first unambiguous
observation of the bound $\Xi^-$ hypernuclear states has been 
reported in the $^{14}$N - $\Xi^-$ system ($^{15}{\!\!\!_{\Xi^-}}$C). 
These authors have performed new analysis of the nuclear emulsion in
the E373 experiment at KEK-PS~\cite{nak15}. The reaction uniquely
identified was  $\Xi^-$ + $^{14}$N $\to $ $^{15}{\!\!\!_{\Xi^-}}$C
$\to$ $^{10}{\!\!\!_{\Lambda}}$Be + $^{5}{\!\!_{\Lambda}}$He. The
$\Xi^-$ - $^{14}$N binding energy (B$_{\Xi^-}$) was measured as 4.43
$\pm$ 0.25 MeV under the assumption that both
$^{10}{\!\!\!_{\Lambda}}$Be and $^{5}{\!\!_{\Lambda}}$He were
produced in their ground states. However, if
$^{10}{\!\!\!_{\Lambda}}$Be was produced in its highest exited
state (with the excitation energy of 3.27 MeV), the B$_{\Xi^-}$ was
1.11 $\pm$ 0.25 MeV. This data is known as Kiso event.

In Ref.~\cite{sun16}, the data of the Kiso event have been analyzed
within the Relativistic mean field (RMF) and Skyrme-Hartree-Fock 
(SHF) models. The stated aim of this work was to determine 
theoretically if the Kiso event corresponds to $^{14}$N - $\Xi^- 
(1s)$ or $^{14}$N - $\Xi^- (1p)$ configurations, and to determine 
the magnitude of the attractive potential that was required to 
reproduce the observed binding energies of the event. However, 
these authors have fitted the effective interactions used in their 
calculations to the measured binding energy of the Kiso event, 
which makes their calculations some what inconsistent.  

The aim of this paper is to study the Kiso event within the QMC 
model. In this model, light quarks within the nonoverlapping 
nucleon and $\Xi^-$ bags (modeled using the MIT bag), interact 
self-consistently with isoscalar-scalar ($\sigma$) and 
isoscalar-vector ($\omega$) mesons in the mean-field approximation
(see, e.g.~\cite{gui88}). The self-consistent response of the bound 
light quarks to the mean $\sigma$ field leads to a new saturation 
mechanism for the nuclear matter~\cite{gui88,gui96,gui06,sai95}. 
The parameters of the QMC model (quark-meson couplings and quark 
masses) were fixed from the nuclear matter  saturation properties 
(see, e.g. Table 13 of Ref.~\cite{sai07}). Using same set of 
parameters, the QMC model has been successfully applied to study 
the properties of finite nuclei~\cite{sai96} and hypernuclei
\cite{tsu98a}, the binding of $\omega$, $\eta$, $\eta^\prime$ and 
$D$ nuclei~\cite{tsu98b,tsu99,kre17} and also the effect 
of the medium on $J/\Psi$ production~\cite{sai07,kre17,sib99}. 
We emphasize that no parameter of the model is fitted to the 
experimental binding of the hypernuclei under study.  

We have used the QMC model~\cite{gui08} to predict the binding energies of 
the states of the $^{15}{\!\!\!_{\Xi^-}}$C hypernucleus having 
the  $1s$ and $1p$ $\Xi^-$ configurations. The corresponding 
scalar and vector mean fields that lead to the binding energies 
predicted by the model are also explicitly calculated. The upper 
and lower components of ${\Xi^-}$ bound state spinors are 
calculated for each configuration, which have been employed to 
calculate the cross sections for the production of the 
$^{15}{\!\!\!_{\Xi^-}}$C hypernuclear spectrum via the $(K^-, 
K^+)$ reaction on a $^{15}$O target within a covariant effective 
Lagrangian model~\cite{shy12,shy17}. This may be useful in 
identifying the states of this hypernucleus more clearly in a 
possible future experiment at the JPARC facility in Japan. 
 
In the Sec. II, our formalism is briefly described, where we have
given some details of the QMC model and the parameters involved
therein. The results of our calculations and their discussions 
are given in Sec. III. The summary and conclusions of our work 
are described in Sec. IV.

\section{Formalism}

In the QMC model~\cite{gui88,gui96}, quarks within the
nonoverlapping nucleon and hyperon bags~\cite{tsu98a} (modeled 
using the MIT bag), interact self-consistently with 
isoscalar-scalar ($\sigma$) and isoscalar-vector ($\omega$) mesons 
in the mean field approximation. Thus, this model self-consistently 
relates the dynamics of the internal quark structure of a hadron to 
the relativistic mean fields arising in the nuclear matter.

In order to calculate the properties of finite hypernuclei, and the
spinors of the bound states, we construct a simple, relativistic shell
model, with the nucleon core calculated in a combination of
self-consistent scalar and vector mean fields. The Lagrangian density
for a hypernuclear system ($Z$) in the QMC model is written as a sum of
two terms, ${\La}^{Z}_{QMC}$ = ${\La^N}_{QMC} + {\La}^{HY}_{QMC}$ ($HY$
= hyperon)~\cite{tsu98a}, where,
\begin{eqnarray}
&&\hspace{-1.4cm}{\La}^N_{QMC} =  \bar{\psi}_N(\vec{r})
[ i \gamma \cdot \partial - M_N(\sigma) - (\, g_\omega \omega(\vec{r}) \nn \\
&& + g_\rho \frac{\tau^N_3}{2}b(\vec{r})
+ \frac{e}{2} (1+\tau^N_3) A(\vec{r}) \,) \gamma_0 ] \psi_N(\vec{r}) \nn \\
%
&& - \frac{1}{2}[ (\nabla \sigma(\vec{r}))^2 + m_{\sigma}^2
      \sigma(\vec{r})^2 ] \nn \\
&& + \frac{1}{2}[ (\nabla \omega(\vec{r}))^2 + m_{\omega}^2 \omega(\vec{r})^2 ]
\nn \\
&& + \frac{1}{2}[ (\nabla b(\vec{r}))^2 + m_{\rho}^2 b(\vec{r})^2 ]
 +\frac{1}{2} (\nabla A(\vec{r}))^2, \label{Lag2}
\end{eqnarray}
and
\begin{eqnarray} 
&&\hspace{-1.4cm}{\La}^{HY}_{QMC} = \sum_{HY=\Lambda,\Sigma,\Xi}
\overline{\psi}_{HY}(\vec{r}) [ i \gamma \cdot \partial - M_{HY}(\sigma)
- (\, g^{HY}_\omega \omega(\vec{r})\nn \\
&& + g^{HY}_\rho I^{HY}_3 b(\vec{r})
+ e Q_{HY} A(\vec{r}) \,) \gamma_0 ] \psi_{HY}(\vec{r}). \qquad \label{Lag3}
\end{eqnarray}
In Eqs.~(\ref{Lag2}) and (\ref{Lag3}), $\psi_N(\vec{r})$, $\psi_{HY}(\vec{r})$, 
$\sigma(\vec{r})$, $\omega(\vec{r})$ and $b(\vec{r})$ are, respectively, the 
nucleon, hyperon, $\sigma$-meson, $\omega$-meson and $\rho$-meson fields, 
while $m_\sigma$, $m_\omega$ and $m_{\rho}$ are the masses of the $\sigma$, 
$\omega$ and $\rho$ mesons. The $A(\vec{r})$ is Coulomb field. $g_\omega$ 
and $g_{\rho}$ are the $\omega$-N and $\rho$-N coupling constants, which 
are related to the corresponding $(u,d)$-quark-$\omega$ ($g_\omega^q$), and 
$(u,d)$-quark-$\rho$ ($g_\rho^q$) coupling constants as $g_\omega = 3 
g_\omega^q$ and $g_\rho = g_\rho^q$. For the hyperon, $g_\omega^{HY} = 
(n_0/3) g_\omega^q$ with $n_0$ being the light-quark number in the hyperon, 
and $g_\rho^{HY} \equiv g_\rho = g_\rho^q$, where $HY$ represents the 
hyperon type. $I^{HY}_3$ and $Q_{HY}$ are the third component of the 
hyperon isospin operator and its electric charge in units of the proton 
charge, $e$, respectively.

The following set of equations of motion are obtained for the hypernuclear
system from the Lagrangian density Eqs.~(\ref{Lag2})-(\ref{Lag3}):
\begin{eqnarray}
&&\hspace{-1.4cm}[i\gamma \cdot \partial -M_N(\sigma)-
(\, g_\omega \omega(\vec{r}) + g_\rho \frac{\tau^N_3}{2} b(\vec{r}) \nn \\
&&\hspace{-1.4cm} + \frac{e}{2} (1+\tau^N_3)
 A(\vec{r}) \,) \gamma_0 ] \psi_N(\vec{r}) =  0, \label{eqdiracn1}
\end{eqnarray}
\begin{eqnarray}
&&\hspace{-1.4cm}[i\gamma \cdot \partial - M_{HY}(\sigma)-
(\, g^{HY}_\omega \omega(\vec{r}) + g^{HY}_\rho I^{HY}_3 b(\vec{r}) \nn \\
&&\hspace{-1.4cm} + e Q_{HY} A(\vec{r}) \,) \gamma_0 ] \psi_{HY}(\vec{r}) = 0, 
\label{eqdiracy2}
\end{eqnarray}
\begin{eqnarray}
&&\hspace{-1.4cm}(-\nabla^2_r+m^2_\sigma)\sigma(\vec{r})  = \nn \\
&&\hspace{-1.4cm} g_\sigma C_N(\sigma) \rho_s(\vec{r})
 + g^{HY}_\sigma C_{HY}(\sigma) \rho^{HY}_s(\vec{r}), \label{eqsigma}
\end{eqnarray}
\begin{eqnarray}
&&\hspace{-1.4cm}(-\nabla^2_r+m^2_\omega) \omega(\vec{r})  =
g_\omega \rho_B(\vec{r}) + g^{HY}_\omega
\rho^{HY}_B(\vec{r}),\label{eqomega}
\end{eqnarray}
\begin{eqnarray}
&&\hspace{-1.4cm}(-\nabla^2_r+m^2_\rho) b(\vec{r})  =
\frac{g_\rho}{2}\rho_3(\vec{r}) + g^{HY}_\rho I^{HY}_3 \rho^{HY}_B(\vec{r}),
 \label{eqrho}
\end{eqnarray}
\begin{eqnarray}
&&\hspace{-1.4cm}(-\nabla^2_r) A(\vec{r})  =
e \rho_p(\vec{r})
+ e Q_{HY} \rho^{HY}_B(\vec{r}),\label{eqcoulomb}
\end{eqnarray}
where, $\rho_s(\vec{r})$ [$\rho^{HY}_s(\vec{r})$], and $\rho_B(\vec{r})$
[$\rho^{HY}_B(\vec{r})$] are, respectively, the scalar, and baryon
densities at position $\vec{r}$ in the hypernucleus, while 
$\rho_3(\vec{r})$ and $\rho_p(\vec{r})$ are, respectively, the third component of 
isovector, and proton densities at the same position in this system
\cite{tsu98a}. On the right hand side of Eq.~(\ref{eqsigma}), a new 
and characteristic feature of QMC model appears that arises from the 
internal structure of the nucleon and hyperon, namely, $g_\sigma 
C_N(\sigma)= - \frac{\partial M_N(\sigma)}{\partial \sigma}$ 
and $g^{HY}_\sigma C_{HY}(\sigma)= - \frac{\partial M_{HY} (\sigma)} 
{\partial \sigma}$ where $g_\sigma \equiv g_\sigma (\sigma=0)$ and 
$g^{HY}_\sigma \equiv g^{HY}_\sigma (\sigma=0)$. We use the nucleon 
and hyperon masses as parameterized in Ref.~\cite{gui08} with the  
corresponding parameters obtained therein. The scalar and vector 
fields as well as the spinors for hyperons and nucleons, can be 
obtained by solving these coupled equations self-consistently.

\begin{figure}[t]
\centering
\includegraphics[width=.50\textwidth]{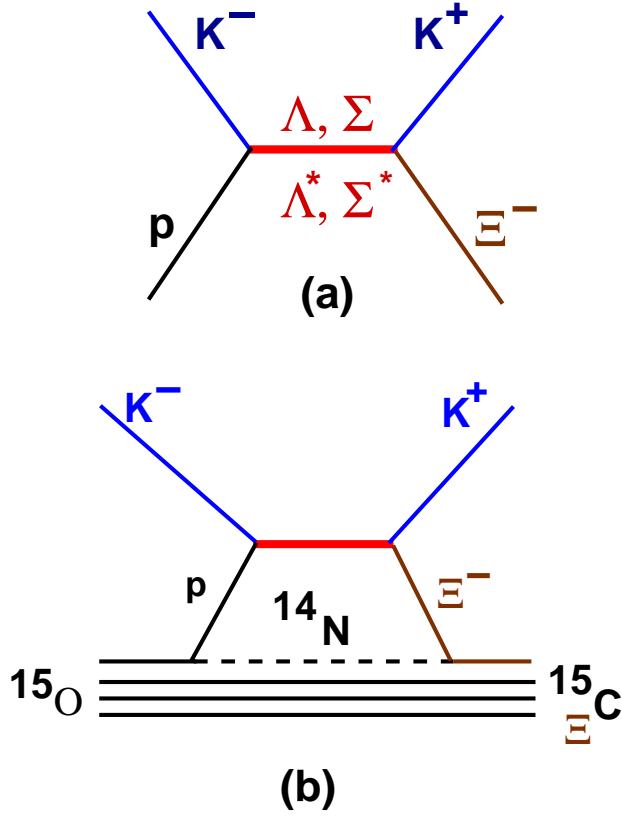}
\caption{(color online)
Graphical representation of our model to describe $p(K^-, K^+)\Xi^-$
(Fig.~1a) and $^{15}$O$(K^-, K^+){_{\Xi^-}}\!\!{^{15}}$C reactions
(Fig. 1b). 
}
\label{fig:Fig1}
\end{figure}

In order to provide further constraint on the ground state structure
of $^{15}{\!\!\!_{\Xi^-}}$C, we have calculated the cross sections
for the production of this hypernucleus via the $(K^-, K^+)$ reaction
on a $^{15}$O target within an effective Lagrangian model (ELM)
\cite{shy12}, which is similar to that used  in Ref.~\cite{shy11} to
study the elementary production reaction, $p(K^-, K^+)\Xi^-$. We
consider only the $s$-channel production diagrams [see, Fig.~1(a)] as
we are interested in the region where $p_{K^-}$ lies below 2 GeV/c.
The model retains the full field theoretic structure of the
interaction vertices and treats baryons as Dirac particles. Figure
1(b) provides the graphical representation of the model used to 
describe the hypernuclear production. The initial state interaction 
of the incoming $K^-$ with a bound target proton leads to excitation 
of intermediate $\Lambda$ and $\Sigma$ resonant states, which 
propagate and subsequently decay into a $\Xi^-$ hyperon that gets 
captured into one of the nuclear orbits, while the other decay 
product, the $K^+$ goes out. In Ref.~\cite{shy11}, it was shown that 
six intermediate resonant states, $\Lambda$, $\Lambda(1405)$, 
$\Lambda(1520)$, $\Lambda(1810)$, $\Sigma$, and $\Sigma(1385)$,  
make the most significant contributions to the cross sections of 
the elementary process. Therefore, like in Ref.~\cite{shy12}, in 
our present study the amplitudes corresponding to these six resonant 
states have been considered.

We assume the initial proton bound state to have quantum numbers of
the outermost {$1p_{1/2}$} proton orbit. The pure
single-hole-single-particle $(p^{-1} \Xi^-)$ configuration has been
considered for the nuclear structure part that leads to the
hypernuclear spectrum that is clearly divided into two groups:
($1p_{1/2}^{-1},1s_{1/2}^{\Xi^-}$), and
($1p_{1/2}^{-1},1p_{3/2}^{\Xi^-}$), corresponding to ($1s_{1/2}$)
and ($1p_{3/2}$) $\Xi^-$ configurations, respectively, for the 
$^{14}$N - $\Xi^-$ system. Since $1p_{3/2}$ proton hole state has a 
much larger binding energy, any configuration mixing is expected to 
be negligible and has not been considered in this study.

In the calculations of the hypernuclear production cross sections with
a particular ground state structure, the momentum space bound
$\Xi^-$ spinors of that state enter into the calculations. In
addition, the spinors for the bound proton in the initial state are
also required. These are the four component Dirac spinors, which are
solutions of the Dirac equation for a bound state problem in the
presence of external potential fields. They are calculated within the
QMC model~\cite{gui08}. The use of bound state spinors determined within 
this model provides an opportunity to investigate the role of the quark 
degrees of freedom in the cascade hypernuclear production.

The proton and $\Xi^-$ spinors are represented by $\psi(k_p)$ and
$\psi(k_{\Xi^-})$, respectively, later on. We write
\begin{eqnarray}
\psi(k_i) & = & \delta(k_{i0}-E_i)\begin{pmatrix}
                    {f(K_i) {\cal Y}_{\ell 1/2 j}^{m_j} (\hat {k}_i)}\\
                    {-ig(K_i){\cal Y}_{\ell^\prime 1/2 j}^{m_j}
                     (\hat {k}_i)} 
                    \end{pmatrix}. \label{spinor}
\end{eqnarray}

In our notation $k_i$ represents a four momentum, and $\vec{k_i}$ a three
momentum. The magnitude of $\vec {k_i}$ is represented by $K_i$, and its
directions by $\hat {k}_i$. $k_{i0}$ represents the timelike component of
momentum $k_i$. In Eq.~(\ref{spinor}), $f(K_i)$ and $g(K_i)$ are the 
radial parts of the upper and lower components of the spinor $\psi(k_i)$ 
with $i$ representing either a proton or a $\Xi^-$. The coupled spherical
harmonics, ${\cal Y}_{\ell 1/2 j}^{m_j}$,  is given by
\begin{eqnarray}
{\cal Y}_{\ell 1/2 j}^{m_j}(\hat {k}_i) & = & <\ell m_\ell 1/2 \mu | j m_j>
                        Y_{\ell m_\ell}(\hat {k}_i) \chi_{\mu},\label{eq.7}
\end{eqnarray}
where $Y_{\ell m_\ell}$ represents the spherical harmonics, and $\chi_{\mu}$
the spin-space wave function of a spin-$\frac{1}{2}$ particle. In
Eq.~(\ref{spinor}) $\ell^\prime = 2j - \ell$ with $\ell$ and $j$ being the
orbital and total angular momenta, respectively.

For the evaluation of the amplitudes corresponding to the processes shown
in Fig.~1(b), one requires the effective Lagrangians at the
resonance-kaon-baryon vertices and the corresponding vertex coupling
constants. In addition, also needed are the propagators for spin-1/2 and
spin-3/2 resonances that connect the initial and final state vertices. They
have all been described in Ref.~\cite{shy12}, and were taken to be the same
in the present study.  Full details of the calculations for the production
cross sections of the $\Xi^-$ hypernuclei within the ELM are given in Ref.
\cite{shy12} and we refer to this article for further details.

\section{Results and discussions}

In Table I, we show the $\Xi^-$ single-particle energies for the 
$^{14}$N nuclear core plus one $\Xi^-$ configuration for the hypernucleus 
$^{15}{\!\!\!_{\Xi^-}}$C as predicted by the QMC model. The listed 
single-particle energy of a particular state can be equated with the $\Xi^-$ 
removal energy (or the binding energy) of the bound $^{14}$N - $\Xi^-$ system 
for that state. 
\begin{table}[t]
\centering
\caption{\label{table1} $\Xi^-$ single-particle energies for a nuclear  
core ($^{14}$N) plus one $\Xi^-$ configuration of the hypernucleus 
$^{15}{\!\!\!_{\Xi^-}}$C for different states. 
}
\vspace{0.2cm}
\begin{tabular}{|c|c|} \hline
 State & single-particle energy  \\
       &(\footnotesize{MeV})     \\
\hline
$^{15}{\!\!\!_{\Xi^-}}$C$(1s_{1/2})$ & -5.366  \\
$^{15}{\!\!\!_{\Xi^-}}$C$(1p_{3/2})$ & -0.779  \\
$^{15}{\!\!\!_{\Xi^-}}$C$(1p_{1/2})$ & -0.099  \\
\hline
\end{tabular}
\end{table}
\noindent
 
We note from this Table that the single-particle energy  of the 
$^{15}{\!\!\!_{\Xi^-}}$C$(1s_{1/2})$ hypernuclear state predicted 
by the QMC model, is within less than 1 MeV of the measured value 
(4.38 $\pm$ 0.25 MeV) of the binding energy of the $^{14}$N - $\Xi^-$ 
system reported in Ref.~\cite{nak15}. This is the case when both of 
the final state nuclei are produced in their ground states in the 
reaction $\Xi^-$ + $^{14}$N $\to$ $^{15}{\!\!\!_{\Xi^-}}$C $\to$ 
$^{10}{\!\!\!_{\Lambda}} $Be + $^{5}{\!\!_{\Lambda}}$He, which are 
identified in the analysis of Ref.~\cite{nak15}. The slight 
overestimate of the experimental data in the QMC calculations 
should be viewed in the light of the fact that we have not tried to 
readjust any of the parameters of the model in order to reproduce 
the experimental binding energies. All the parameters have been 
held fixed to those given in Ref.~\cite{gui08} and used in all the 
previous calculations of the $\Xi^-$ hypernuclei~\cite{shy12,shy13,
tsu98a} within the QMC model. Furthermore, with the same set of 
parameters the single-particle energy of the  $1s_{1/2}$ state of the 
$^{12}{\!\!\!_{\Xi^-}}$Be hypernucleus is predicted to be -3.04 MeV, 
which is closer to the values reported in several other theoretical 
approaches~\cite{mat11,hiy08}.

The single-particle energy of the $^{15}{\!\!\!_{\Xi^-}}$C$(1p_{3/2})$
state is -0.779 MeV, which is closer to the lower limit of the binding 
energy of the $^{14}$N - $\Xi^-$ system (1.11 $\pm$ 0.25 MeV) if  
$^{10}{\!\!\!_{\Lambda}}$Be is produced in its highest exited state 
(which is supposed to be 3.27 MeV as predicted in cluster model 
calculations of Ref.~\cite{hiy08}). On the other hand, the 
$^{15}{\!\!\!_{\Xi^-}}$C$(1p_{1/2})$ state is almost unbound. Thus as 
per QMC model, the state of the $^{14}$N - $\Xi^- $ system observed in 
Ref.~\cite{nak15} with a binding energy of 1.11 $\pm$ 0.25 MeV, is 
most likely to have the quantum number $1p_{3/2}$.

The scalar and vector mean fields calculated self-consistently within 
the QMC model that lead to these states are shown in Fig.~2 (a). It 
may be recalled that in this model the scalar and vector fields are 
generated by the couplings of $\sigma$ and $\omega$ mesons to light 
quarks. Because of the different masses of these mesons and their 
couplings to the light-quark fields the scalar and vector fields can 
acquire different radial dependence. We note that both the scalar and 
vector QMC fields have their maxima away from the point $r = 0$ (center 
of the hypernucleus). In the mean field models of the finite nuclei the 
proton densities are somewhat pushed out as compared to those of the 
neutron, because of the Coulomb repulsion. In fact, the proton density 
distribution has a peak around 2 fm away from the center in 
$^{15}{\!\!\!_{\Xi^-}}$C. This causes the $\Xi^-$ potential to have a 
peak around 2 fm outside of the center of the nucleus, as a consequence 
of the self-consistent procedure of including the Coulomb force. It may 
be noted that the vector potential felt by $\Xi^-$ contains also the 
attractive Coulomb potential

\begin{figure}[!t]
\begin{center}
\includegraphics[scale=.50]{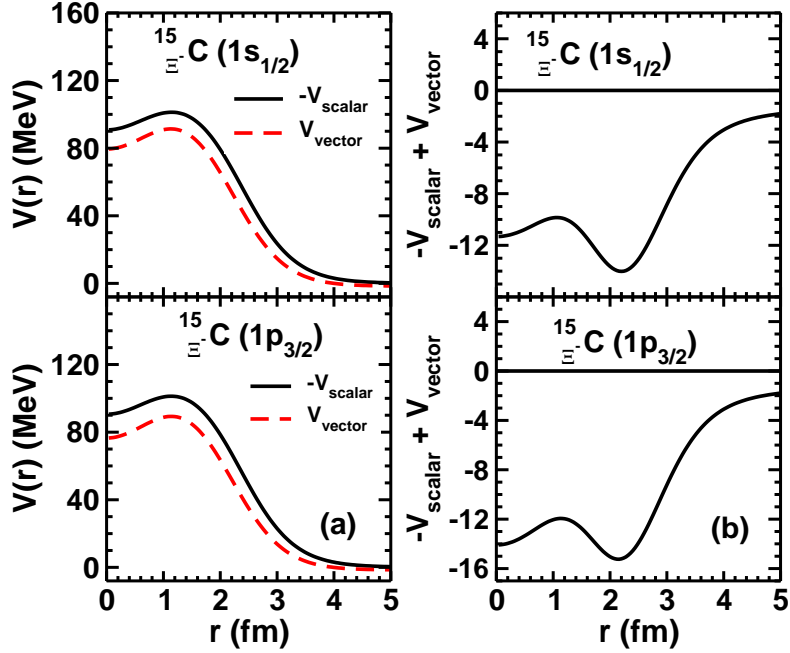}
\end{center}
\caption{(color online)
(a)~Vector (V$_{vector}$) and scalar (V$_{scalar}$) potential fields 
felt by $\Xi^-$ in $1s_{1/2}$, and $1p_{3/2}$ states in the hypernucleus 
$^{15}{\!\!\!_{\Xi^-}}$C as a function of the radial distance $r$ from 
the center of the hypernucleus. (b)~The sum -V$_{scalar}$ + V$_{vector}$ 
as a function of $r$. 
}
\label{Fig2}
\end{figure}

In Fig.~2 (b), we show the quantity -V$_{scalar}$ + V$_{vector}$ (V$_{NR}$)
as a function of the radial distance $r$ from the center of the nucleus. 
V$_{NR}$ should be compared with the corresponding Woods-Saxon $\Xi^-$ - 
nucleus potential of a non-relativistic theory~\cite{shy12,shy13}. We note 
that at $r = 0$, the value of V$_{NR}$ is about -11 MeV for the $s$-state and 
-14 MeV for the $p$ state. This is similar to the depth of the $\Xi^-$ - 
$^{12}$C potential assumed in Ref.~\cite{kha00} in order to interpret the 
observed spectrum of the $(K^-,K^+)$ reaction on a $^{12}$C target. 
Interestingly, V$_{NR}$ has a dip around $r = 2$ fm. This is the consequence 
of the presence of an attractive Coulomb term in V$_{vector}$, due to the 
pushed out proton density distribution from the center of the nucleus, 
as already explained.

We next discuss the production of the $^{15}{\!\!\!_{\Xi^-}}$C hypernucleus
in the $(K^-,K^+)$ reaction on  a $^{15}$O target.  For the evaluation of 
the amplitudes corresponding to the processes shown in Fig.~1(b), one 
requires the effective Lagrangians at the resonance-kaon-baryon vertices 
and the corresponding vertex coupling constants. In addition, the 
propagators for spin-1/2 and spin-3/2 resonances that connect the initial 
and final state vertices, are also needed. They have all been taken from 
the Ref.~\cite{shy12}. Furthermore, the spinors for the bound proton in the 
initial state as well as for the bound $\Xi^-$ in the final state also 
enter in these calculations. They have been determined within the improved 
QMC model~\cite{gui08}.  
\begin{figure}[!t]
\includegraphics[scale=.50]{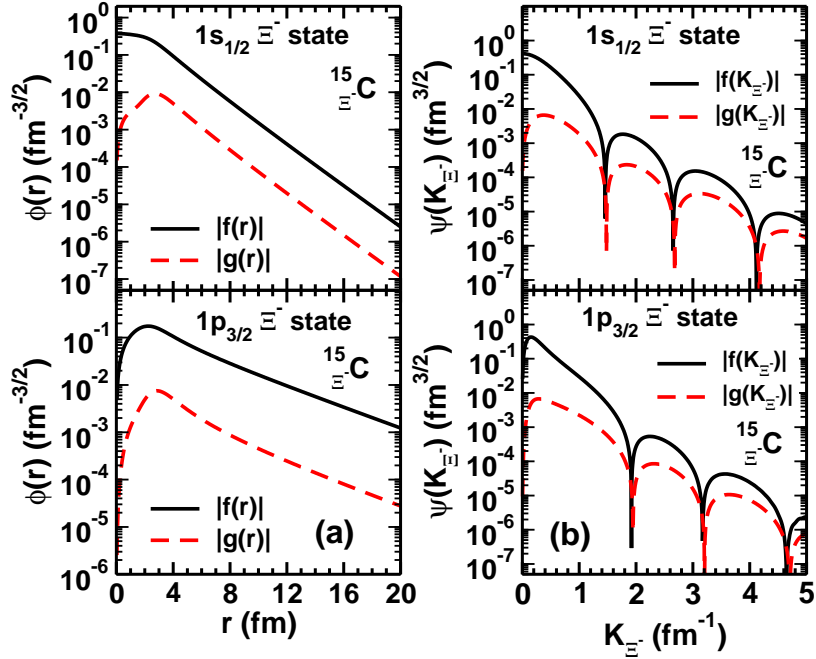} 
\caption{(color online)
(a) Moduli of the upper ($|f(r)|$) and lower ($|g(r)|$) components of the
coordinate space spinors for 1$s_{1/2}$ and 1$p_{3/2}$ $\Xi^-$ states in  
$^{15}_{\Xi^-}$C. Solid lines represent the upper component while the 
dashed line the lower component. (b) Moduli of upper (solid lines)  and 
lower (dashed lines) components of the momentum space spinors of the 
$\Xi^-$ bound states in  $^{15}_{\Xi^-}$C for the same states as in (a).
}
\label{fig:Fig3}
\end{figure}

Figs.~3(a) and 3(b) show the moduli of the upper and lower components of
$1s_{1/2}$, and $1p_{3/2}$ $\Xi^-$ spinors for the $^{15}_{\Xi^-}$C 
hypernucleus in coordinate space and momentum space, respectively. 
The spinors in the momentum space are obtained by Fourier transformation 
of the corresponding coordinate space spinors.  We  note that only for $K
$ values below 1.5 fm$^{-1}$, are the magnitudes of the lower components,
$|g(K)|$, substantially smaller than those of the upper components. In 
the region of $K$ pertinent to the cascade hypernuclear production, $
|g(K)|$ may not be negligible. Thus the relativistic effects resulting 
from the small component of bound states spinors could be large for the 
hypernuclear production reactions on nuclei (see also, Ref.~\cite{ben89}).

In Fig.~4, we study the $0^\circ$ differential cross sections  
$[(d\sigma/d\Omega)^{0^\circ}]$ for the reaction $K^- + ^{15}$O $\to K^+ 
+ {^{15} {\!\!\!_{\Xi^-}}}$C obtained by using the proton-hole and 
$\Xi^-$ bound state spinors calculated within the QMC model. The 
threshold beam momentum for this reaction is about 0.756 GeV/c, whereas 
that for the elementary reaction $K^- + p \to K^+ + \Xi^-$, is about 1.0 
GeV/c. The $\Xi^-$ hyperon in a $1s_{1/2}$ state can populate 0$^-$ and 
1$^-$ states of the hypernuclear spectrum corresponding to the 
hole-particle configuration $[(p_{1p_{1/2}}^{-1},\Xi^-_{1s_{1/2}})]$, 
while that in a $1p_{3/2}$ state can populate 1$^+$, and 2$^+$ states 
according to the configuration $[(p_{1p_{1/2}}^{-1},\Xi^-_{1p_{3/2}})]$. 
However, the states of unnatural parity are very weakly excited due to 
vanishingly small spin-flip amplitudes for this reaction~\cite{gal83}.

In Figs.~4(a) and 4(b), $(d\sigma/d\Omega)^{0^\circ}$ for this reaction 
are shown as a function of beam momentum (p$_{K^-}$) for populating the 
hypernuclear states 1$^-$ and 2$^+$, respectively. These are states of 
maximum spin with natural parity corresponding to the cases where 
$\Xi^-$ hyperon is in $1s_{1/2}$ and $1p_{3/2}$ orbits, respectively. 
It is established from the studies reported in Refs.~\cite{shy12,shy17} 
that within a given hole-particle configuration, states with higher 
angular momenta $J$ have the larger cross sections. The configurations 
of the final hypernuclear states are as described in the figure captions. 
As stated earlier, the relative motions of $K^-$ and $K^+$ mesons in the 
initial and final channels, respectively, are described by the plane 
waves.  However, the distortion effects are approximately accounted for 
by introducing a reduction factor of 3.2 to the overall cross section, 
which is estimated in an eikonal-approximation based procedure as 
discussed in Refs.~\cite{shy06,shy16}. This method primarily takes care 
of the absorption effects in the incoming and outgoing channels. 
\begin{figure}
\centering
\includegraphics[width=.40\textwidth]{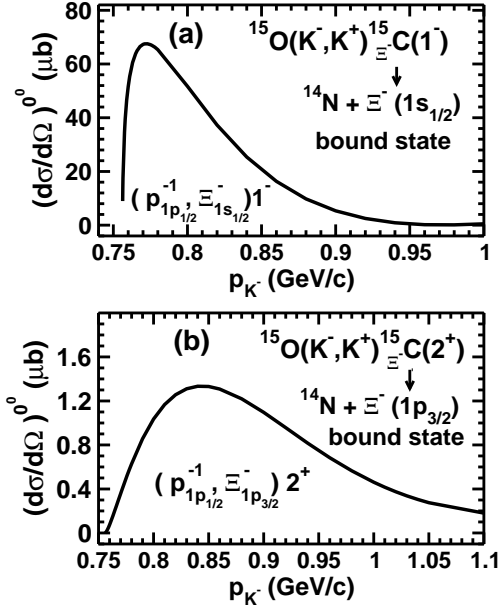}
\caption{(a) Calculated differential cross section at $0^\circ$ as a 
function of $K^-$ beam momentum for the 
$^{15}$O$(K^-,K^+)^{15}{\!\!\!_{\Xi^-}}$C reaction leading to the 
population of the 1$^-$ hypernuclear state within the hole-particle 
configuration $(p_{1p_{1/2}}^{-1}, \Xi^-_{1s_{1/2}})$. These results 
are obtained with the QMC proton and $\Xi^-$ bound state spinors. 
(b) The same as in (a) for populating the 2$^+$ hypernuclear state 
within hole-particle configuration $(p_{1p_{1/2}}^{-1},
\Xi^-_{1p_{3/2}})$. 
}
\label{fig:Fig4}
\end{figure}

In Fig.~4(a) we note that cross section rises sharply as the beam 
momentum increases above the reaction threshold. It peaks around 
p$_{K^-}$ $\approx 0.78$ GeV/c and falls off gradually as p$_{K^-}$ 
increases further. This feature is reminiscent of the results obtained 
for the $^{12}$C$(K^-,K^+)^{12}{\!\!\!_{\Xi^-}}$C $(1^-)$ reaction as
 shown in Ref.~\cite{shy12}. This is also similar to trends seen in 
the corresponding cross section of the elementary reaction 
p$(K^-,K^+)\Xi^-$ shown in Ref.~\cite{shy12}.  In both the cases the 
beam momentum distributions of the cross sections are relatively 
narrow. 

In contrast, the beam momentum distribution of the cross sections 
for populating the $2^+$ state in the $^{15}$O$(K^-,K^+)^{15}
{\!\!\!_{\Xi^-}}$C reaction is comparatively broad as can be 
seen in Fig.~4(b). In this case the cross section peaks around
p$_{K^-}$ $\approx 0.84$ GeV/c, which is about 0.06 GeV/c higher
than that of Fig.~4(a). Also the absolute magnitudes of the 
cross sections are smaller in this case. This behavior has its 
origin in the differences in the momentum space spinors of the 
two states as a function of $K_{\Xi^-}$ as shown in Fig.~3(b). 
The momentum transfers ($K_{trans}$) involved at 0$^\circ$, vary 
between 2.30 - 3.55 fm$^{-1}$. We note that the peaks and dips 
of the spinors as a function of $K_{\Xi^-}$ of the two states 
do not occur at the same position. For example, at the peak 
positions of the cross sections in two cases (where $K_{trans}$ 
$\approx$ 3.00 fm$^{-1}$), the $|f(K_{\Xi^-})|$ for the $1s_{1/2}
$ $\Xi^-$ state is about an order of magnitude larger than that 
of the $1p_{3/2}$ state. The major reason for the different 
behavior of $|f(K_{\Xi^-})|$ for the two states is the difference 
in the corresponding binding energies.

The different behavior of the cross sections seen in Figs.~4(a) 
and 4(b) could be exploited to identify the states of the 
$^{15}{\!\!\!_{\Xi^-}}$C in a possible future experiment on the 
$(K^-,K^+)$ reaction on a $^{15}$O target. 

\section{Summary and conclusions}

In this paper, we studied the structure of the hypernucleus 
$^{15}{\!\!\!_{\Xi^-}}$C (bound state of the $^{14}$N + $\Xi^-$ 
system) within the framework of the quark-meson coupling (QMC) model. 
This work is motivated by the recent observation (Kiso event) that 
provides the first clear evidence for a strongly bound $\Xi^-$ 
hypernuclear state. Recently, in Ref.~\cite{sun16} an analysis of 
this data has been performed  within the relativistic mean field 
and Skyrme-Hartree-Fock models, where the required effective 
interactions were fitted to the experimental binding energy of 
$^{15} {\!\!\!_{\Xi^-}}$C. We do not consider this procedure to 
be fully consistent.   

In contrast to this, none of the parameters of the QMC model 
(quark-meson coupling constants and quark masses) is fitted to the 
experimental data on the binding energy of the hypernucleus under 
investigation. In fact, in our present calculations the values of 
all the model parameters were taken to be the same as those fixed in 
Ref.~\cite{gui08} from the nuclear matter properties. 

Our calculations suggest that for the Kiso event, the observed bound 
state of $^{15}{\!\!\!_{\Xi^-}}$C with a binding energy $B_{\Xi^-} 
\approx$ 4.4 MeV is the ground state of this hypernucleus  having 
a configuration $^{14}$N + $\Xi^-(1s_{1/2})$. The second observed
state with $B_{\Xi^-} \approx$ 1.1 MeV has a configuration $^{14}$N 
+ $\Xi^- (1p_{3/2})$. This could be interpreted as an excited state 
of $^{15}{\!\!\!_{\Xi^-}}$C hypernucleus.

To confirm these interpretations of the Kiso event, we suggest to
perform measurements on the production of $^{15}{\!\!\!_{\Xi^-}}$C
hypernucleus via the $(K^-, K^+)$ reaction on a $^{15}$O target. We 
have calculated the cross sections of this reaction within an 
effective Lagrangian model, using the proton hole and $\Xi^-$ bound 
state spinors of $s$ and $p$ states derived from the same QMC model.  
We have considered the excitation of altogether six $\Lambda$ and 
$\Sigma$ hyperon resonance intermediate states in the initial 
collision of the $K^-$ meson with a target proton. These states 
subsequently propagate and decay into a $\Xi^-$ hyperon and a $K^+$ 
meson. The hyperon gets captured in one of the nuclear orbits, while 
the  $K^+$ meson goes out. We find that $\Xi^-$ hyperon in 
$1s_{1/2}$ and $1p_{3/2}$ states lead to remarkably different 
beam momentum distributions of the zero degree differential 
cross sections. This may help in distinguishing between the two 
states of this hypernucleus.  

\section{Acknowledgments}
The work of one of the authors (RS) is supported by the Science and 
Engineering Research Board (SERB), Department of Science and 
Technology, Government of India under Grant no. SB/S2/HEP-024/2013, 
while the work of the other author (KT) is supported by the Conselho 
Nacional de Desenvolvimento Cient\'{i}fico e Tecnol\'{o}gico - CNPq,   
Process No.~308088/2015-8, and part of the projects, Instituto Nacional de 
Ci\^{e}ncia e Tecnologia - Nuclear Physics and Applications 
(INCT-FNA), Brazil, Process No. 464898/2014-5, and FAPESP Tem\'{a}tico, 
Brazil, Process No. 2017/05660-0.

\end{document}